\pdfoutput=1
\documentclass[aip, rsi, reprint]{revtex4-1}

\usepackage[pdftex, final]{graphicx}
\usepackage{amsmath}
\usepackage{csquotes}
\usepackage{units}

\usepackage{setspace}
\usepackage{color}
\newcounter{NBc}

\begin{document}

\title{Commissioning of the PRIOR proton microscope}

\newcommand{\affGSI} {\affiliation{GSI Helmholtzzentrum f\"ur Schwerionenforschung GmbH,  Darmstadt, Germany}}
\newcommand{\affTECH}{\affiliation{Physics Department, Technion, Haifa, Israel}}
\newcommand{\affITEP}{\affiliation{Institute for Theoretical and Experimental Physics,  Moscow, Russia}}
\newcommand{\affLANL}{\affiliation{Los Alamos National Laboratory, Los Alamos, USA}}
\newcommand{\affTUD} {\affiliation{Technische Universit\"at Darmstadt, Darmstadt, Germany}}
\newcommand{\affIPCP}{\affiliation{Institute of Problems of Chemical Physics, Chernogolovka, Russia}}

\author{D.~Varentsov} \email{d.varentsov@gsi.de}          \affGSI
\author{O.~Antonov}             \affTECH
\author{A.~Bakhmutova}          \affITEP
\author{C.~W.~Barnes}           \affLANL
\author{A.~Bogdanov}            \affITEP
\author{C.R.~Danly}             \affLANL
\author{S.~Efimov}              \affTECH
\author{M.~Endres}              \affTUD
\author{A.~Fertman} \thanks{Presently at \enquote{Skolkovo} Foundation, Russia} \affITEP
\author{A.A.~Golubev}           \affITEP
\author{D.H.H.~Hoffmann}        \affTUD
\author{B.~Ionita}              \affGSI
\author{A.~Kantsyrev}           \affITEP
\author{Ya.E.~Krasik}           \affTECH
\author{P.M.~Lang} \thanks{Presently at European XFEL GmbH, Hamburg, Germany} \affTUD
\author{I.~Lomonosov}           \affIPCP
\author{F.G.~Mariam}            \affLANL
\author{N.~Markov}              \affITEP
\author{F.E.~Merrill}           \affLANL
\author{V.B.~Mintsev}           \affIPCP
\author{D.~Nikolaev}            \affIPCP
\author{V.~Panyushkin}          \affITEP
\author{M.~Rodionova}           \affGSI \affTUD
\author{M.~Schanz}              \affTUD
\author{K.~Schoenberg}          \affLANL
\author{A.~Semennikov}          \affITEP
\author{L.~Shestov}             \affGSI \affTUD
\author{V.S.~Skachkov}          \affITEP
\author{V.~Turtikov} \thanks{Presently at \enquote{Skolkovo} Foundation, Russia} \affITEP
\author{S.~Udrea}   \thanks{Presently at Goethe-Universit\"at Frankfurt am Main, Germany} \affTUD
\author{O.~Vasylyev}            \affGSI
\author{K.~Weyrich}             \affGSI
\author{C.~Wilde}               \affLANL
\author{A.~Zubareva}            \affIPCP

\date{\today}

\begin{abstract}

Recently a new high energy proton microscopy facility PRIOR (Proton Microscope for FAIR) has been designed, constructed and successfully commissioned at GSI Helmholtzzentrum f\"ur Schwerionenforschung (Darmstadt, Germany). As a result of the experiments with \unit[3.5~-- 4.5]{GeV} proton beams delivered by the heavy ion synchrotron SIS-18 of GSI, $30\,\rm \mu m$ spatial and \unit[10]{ns} temporal resolutions of the proton microscope have been demostrated. A new pulsed power setup for studying properties of matter under extremes has been developed for the dynamic commissioning of the PRIOR facility. This paper describes the PRIOR setup as well as the results of the first static and dynamic proton radiography experiments performed at GSI.

\end{abstract}

\maketitle

\section{Introduction}
\label{sec:intro}

Proton radiography or microscopy is a powerful technique for probing the interior of dense objects in dynamic experiments by mono-energetic beams of GeV-energy protons, using a special system of magnetic lenses for imaging and aberrations correction~\cite{mor:2013}. With this technique, one can measure the areal density of a thick sample with sub-percent accuracy, micrometer-level spatial resolution and nanosecond temporal scale. Proton radiography with magnetic lenses was invented in the 1990's at Los Alamos National Laboratory (LANL) as a diagnostic to study dynamic material properties under extreme pressures, densities and strain rates~\cite{kin:1999}. Since that time proton radiography and microscopy facilities have been also commissioned at the Institute for Theoretical and Experimental Physics (ITEP)~\cite{gol:2008,kan:2014} and at the Institute for High Energy Physics (IHEP)~\cite{ant:2010,ant:2013a,mak:2014} in Russia.

The capability of radiographic imaging of dynamic systems with unprecedented spatial, temporal and density resolution is of considerable interest for plasma physics and materials research. Therefore high energy proton microscopy (HEPM) is seen as a key diagnostic for high energy density physics experiments with intense heavy ion and proton beams, which are planned at the future Facility for Anti-proton and Ion Research (FAIR) in Darmstadt, Germany~\cite{sha:2014}. The worldwide unique facility called PRIOR (\textit{Proton Microscope for FAIR}) will employ high-energy (\unit[2 -- 5]{GeV}), high-intensity (up to \unit[$2.5\cdot 10^{13}$]{protons per pulse}) proton beams from the SIS-100 synchrotron at FAIR for multidisciplinary research such as experiments on fundamental properties of materials in extreme dynamic environments generated by different drivers (pulsed power generators, high-energy lasers, gas guns or explosive-driven generators) prominent for warm dense matter research and high energy density physics as well as the PaNTERA (\textit{Proton Therapy and Radiography}) experiment~\cite{dur:2012,var:2013,pra:2015,pra:2015a} for biophysics and medicine. This paper describes the PRIOR setup which has been recently installed and commissioned at the GSI Helmholtzzentrum f\"ur Schwerionenforschung (Darmstadt, Germany) using \unit[3.5~-- 4.5]{GeV} proton beams delivered by the SIS-18 heavy ion synchrotron as well as recent static and dynamic experiments performed with this new HEPM facility.

\section{Design and construction}
\label{sec:constr}

The ion-optical design of the PRIOR microscope~\cite{mer:2009} is analogous to the design of the \unit[800]{MeV} x7 and x3 magnifying lenses at LANL~\cite{mot:2003,mer:2011}, but for proton energies up to \unit[4.5]{GeV}. The PRIOR magnifier employs high-gradient (\unit[120]{T/m}) NdFeB axially and radially segmented permanent magnet quadrupole (PMQ) lenses~\cite{ger:1980, hal:1980, kan:2015} and provides a magnification of about four with a field of view of \unit[15]{mm}. The illuminating proton beam is matched to the magnifier by five upstream quadrupole electromagnets in order to cancel the second-order position-dependent chromatic aberrations of the system~\cite{mot:1997}.

\begin{figure}[ht]
   \includegraphics[width=\columnwidth]{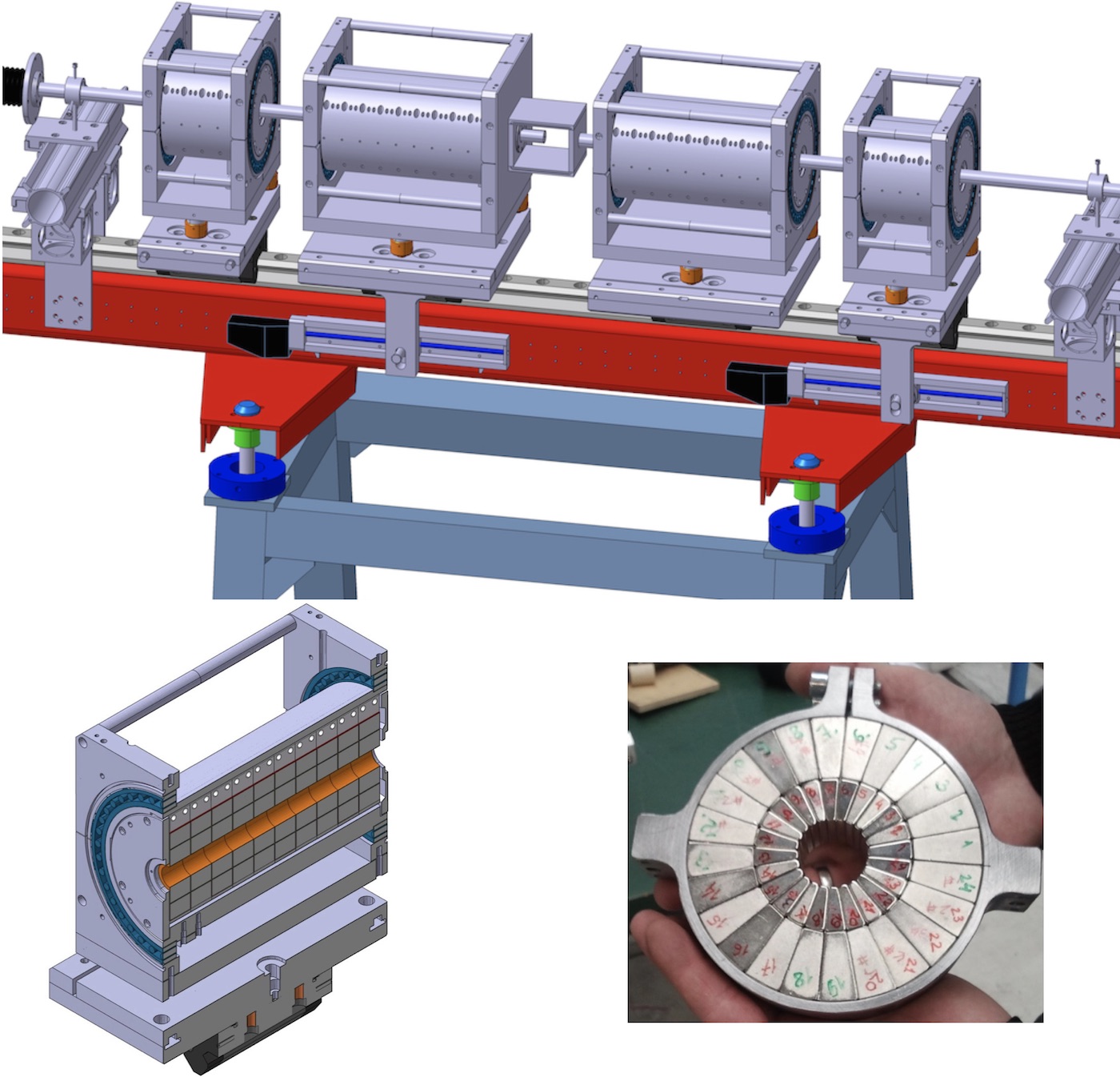}
   \caption{Mechanical design of the PRIOR magnifier: four permanent magnet quadrupole (PMQ) lenses are installed on a common rail. A vacuum collimator box is attached to the third lens. Each quadrupole lens (bottom left) contains up to five or ten \unit[36]{mm}-long modules; each module (bottom right) is built as a double-layer PMQ with \unit[1.8]{T} pole tip field~\cite{kan:2015}. \label{fig:magdraw}}
\end{figure}

The mechanical design of the magnifier and its PMQ lenses is shown in Fig.~\ref{fig:magdraw}. Four quadrupole magnets in a Russian quadruplet configuration~\cite{dym:1963} are installed on a common rail. The aperture of the magnets is \unit[30]{mm}. The outer lenses have a length of \unit[14.4]{cm} and the inner lenses are twice this length, as required by the beam optics. Each magnet is installed on a motorized support table and can be independently moved along the rail for focusing. Each magnet is made out of \unit[36]{mm}-long modules (see Fig.~\ref{fig:magdraw}) which can be individually adjusted in $x,$ $y$ and $\theta$ directions for aligning the magnetic field axes and the mid-plane tilts. With the help of a precision cylindrical field scanner~\cite{lan:2015, ska:1993, gol:2010}, the magnetic axes and the field mid-planes of each lens have been aligned to the accuracy of $\pm 20\,\rm \mu m$ and $\pm 0.1^\circ ,$ respectively~\cite{lan:2015}. Using the same scanner and the field reconstruction procedure~\cite{ska:1993, gol:2010}, which has been used at ITEP for over two decades, 3D field maps of the lenses have been obtained. Similar techniques have been applied for accurate 3D magnetic field description and implemented in the ion-optical calculations~\cite{tak:2013}. A \unit[10]{cm} long tungsten collimator with an elliptical aperture was installed in the Fourier plane of the magnifier (see Fig.~\ref{fig:magdraw}). In order to adjust the contrast of the proton radiographs for a particular target, collimators with angular acceptances $\theta_c$ of 2, 3 and \unit[4]{mrad} were used during the experiments.

\begin{figure}[ht]
   \includegraphics[width=\columnwidth]{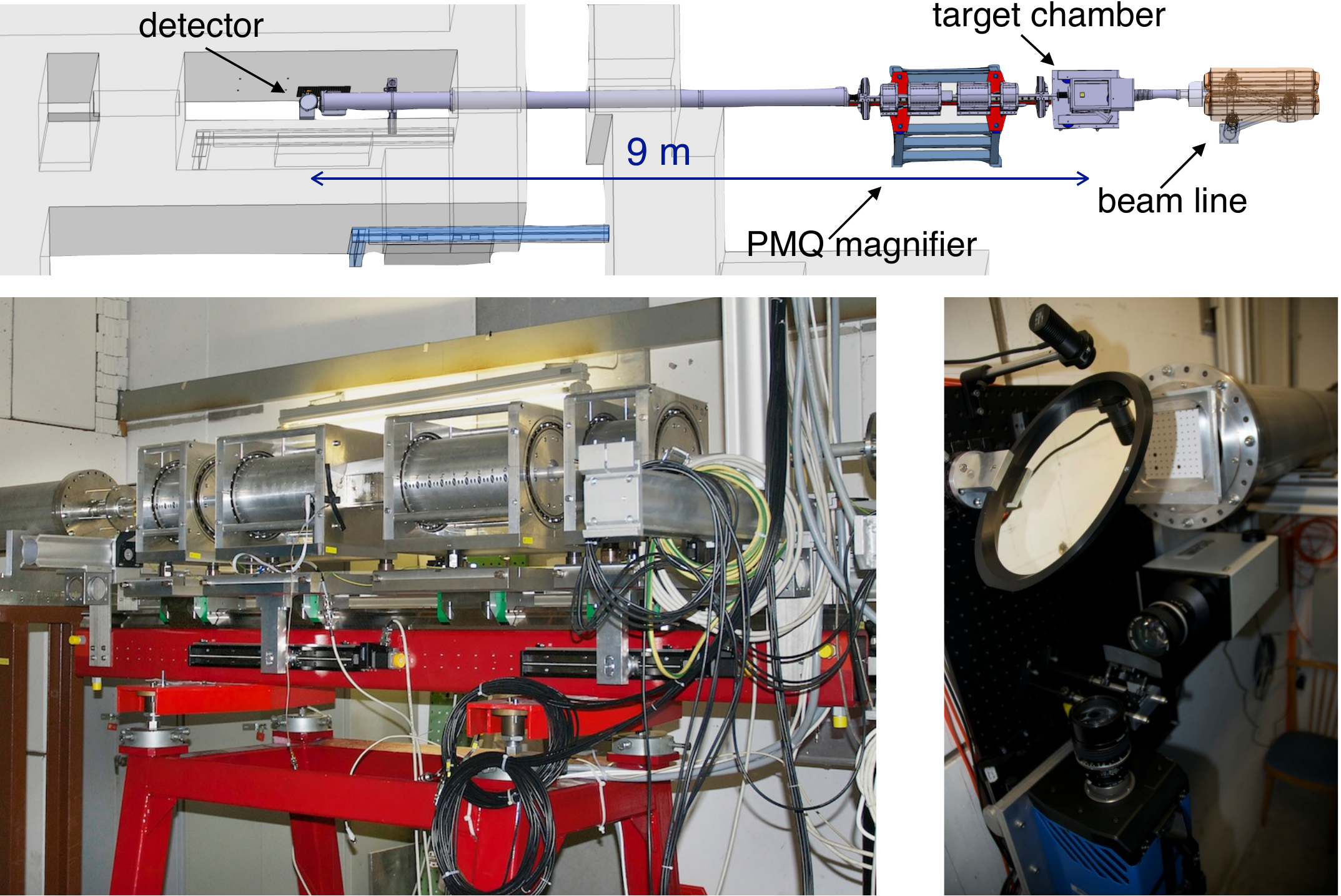}
   \caption{Layout of the PRIOR setup at the HHT experimental area (top), the PMQ magnifier (bottom left) and the detector system (bottom right). \label{fig:layout}}
\end{figure}

Although the PMQ magnifier itself has a length of \unit[1.4]{m}, a long drift after the magnifier is needed to achieve the required magnification (see Fig.~\ref{fig:layout}). Therefore the detector (image collection) system was installed in the newly constructed concrete-shielded detector hall about \unit[9]{m} downstream of the target location. With a pellicle / mirror arrangement, the system employs two cameras simultaneously: a high resolution (4~Mp) CMOS camera (PCO DIMAX HS) used mainly for static experiments and a fast intensified CCD camera (PCO DICAM PRO) for dynamic measurements. \unit[$10\times 10$]{cm} columnar CsI and plastic BC-400 scintillators were installed for static and for dynamic measurements, respectively. Preliminary experiments with \unit[800]{MeV} protons performed at LANL have proven that neither CsI nor plastic scintillators show any image quality or light output degradation even for large irradiation doses ($\approx10^{11}$ protons per $\rm cm^2$)~\cite{lan:2015}.

\section{Static experiments}
\label{sec:static}

The PRIOR setup has been commissioned using \unit[3.5~-- 4.5]{GeV} proton beams from the SIS-18 synchrotron of GSI using only moderate intensity ($10^8$ protons per pulse) beams for the static experiments with \unit[3.6]{GeV} protons. A proton transmission image (radiograph) is obtained as the ratio between the raw images taken with and without an object present under the same proton beam conditions~\cite{kin:1999}. For this purpose a \enquote{beam image} (image without an object) is always recorded shortly before or after imaging an object. The distribution of the areal density is then obtained by applying a transmission~-- density calibration (see Eq.~\eqref{eq:dens} below). Due to the rather low beam intensity and the shot-to-shot beam position fluctuations, in order to enhance the contrast and flatten the background of the radiographs, about 20~-- 50 target and beam images were recorded and averaged in the PRIOR static commissioning experiments.

For tuning and measuring the performance of the PRIOR microscope, a large set of small static test objects were prepared. The targets were placed in a vacuum target chamber equipped with a precision 6-axis manipulator. The whole setup was evacuated to the $10^{-3}\,\rm mbar$ pressure in order to maintain radiographic resolution.

\begin{figure}[ht]
   \includegraphics[width=\columnwidth]{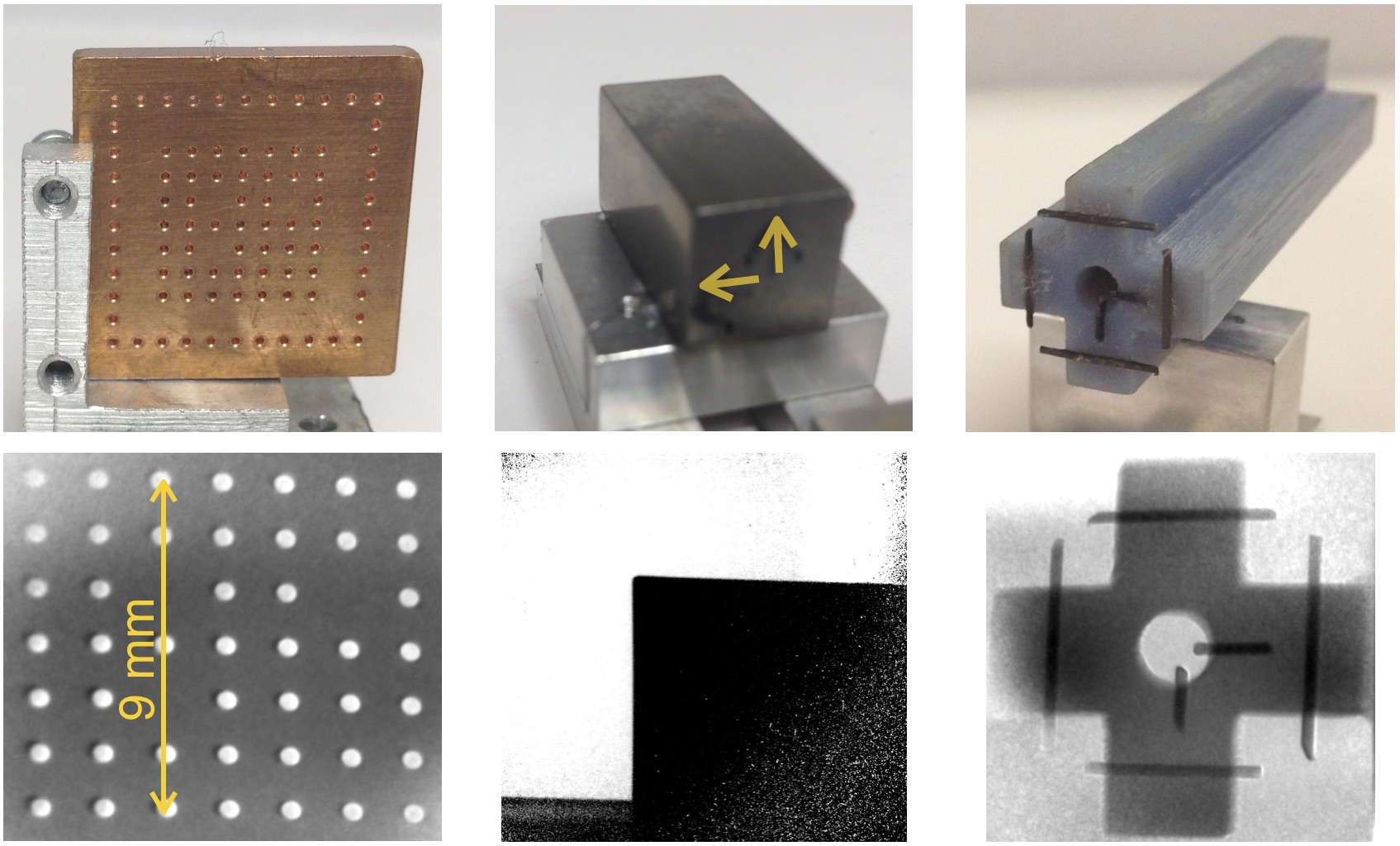}
   \caption{Targets for static commissioning of the PRIOR microscope (top) and their proton radiographs (bottom): \enquote{fiducial plate} for image tuning and spatial calibration (left), \enquote{rolled edge} for spatial resolution measurements (middle) and \enquote{Maltese cross} for verification of the matching conditions. \label{fig:static}}
\end{figure}

The most utilized targets and their proton radiographs are shown in Fig.~\ref{fig:static}. The \enquote{fiducial plate} is a \unit[3]{mm} thick copper plate with \unit[0.5]{mm} holes machined at \unit[1.5]{mm} spacing. It was used as a relatively thin target for quick tuning and controlling image distortions as well as for providing spatial calibration while minimizing activation and radiation in the experimental area.

\begin{figure}[ht]
   \includegraphics[width=\columnwidth]{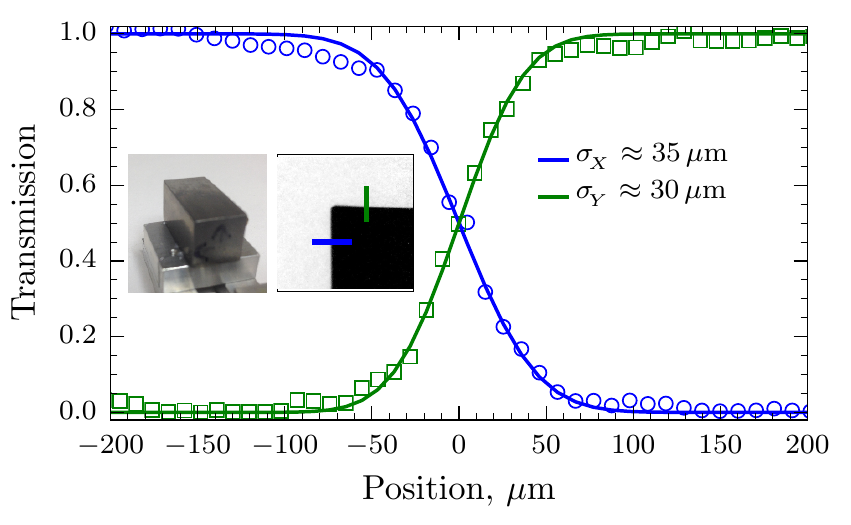}
   \caption{Spatial resolution measurements with rolled edge. Measured horizontal (blue) and vertical (green) edge transitions near the center of the image are shown along with the corresponding error function fits. The horizontal and vertical RMS widths of this edge were measured to be $35\,\rm\mu m$ and $30\,\rm\mu m,$ respectively. \label{fig:edge}}
\end{figure}

The spatial resolution of the microscope has been measured using the \enquote{rolled edge} target~--- a \unit[20]{mm}-thick tungsten slab. Two sides of this edge (marked by arrows in Fig.~\ref{fig:static}) are rolled with a \unit[500]{mm} radius which makes the measurements insensitive to beam-target tilt misalignments of a few milliradian. The spatial resolution can be defined as the standard deviation of the derivative of the measured edge transition (line spread function, LSF). To accurately determine resolution one also needs to deconvolve the known width of the rolled edge from the measured density profile to get the blur function. However since the extent of the rolled edge itself is only a small contribution ($\sigma_{\!\scriptscriptstyle\rm edge} \approx 5\,\rm\mu m$), for tuning and quick analysis we have fit the edge transition to an error function which provides a good estimate of the LSF root mean square (RMS) width. Figure~\ref{fig:edge} shows the horizontal and vertical edge transmission profiles along with the error function fits. These fits resulted in a horizontal resolution of $\sigma_x = 35\,\rm \mu m$ and a vertical resolution of $\sigma_y = 30\,\rm \mu m.$

The \enquote{Maltese cross} target (Fig.~\ref{fig:static}, right) was used to check the matching conditions~\cite{mot:1997} required for canceling the position-dependent second-order chromatic aberrations of the microscope. The target is an elongated piece of plastic with Maltese cross like shape and \unit[0.5]{mm} diameter tungsten wires glued on its back side. When an image of the wires is formed by the magnifier, the protons which were penetrating through both the plastic and the wires have a smaller energy than those which saw only the wires. If the proton beam is not properly matched to the microscope in X- or Y-plane, this difference in the energy loss will result in a slight shifting of the corresponding wires images at the Maltese cross boundaries.

\begin{figure}[ht!]
   \includegraphics[width=\columnwidth]{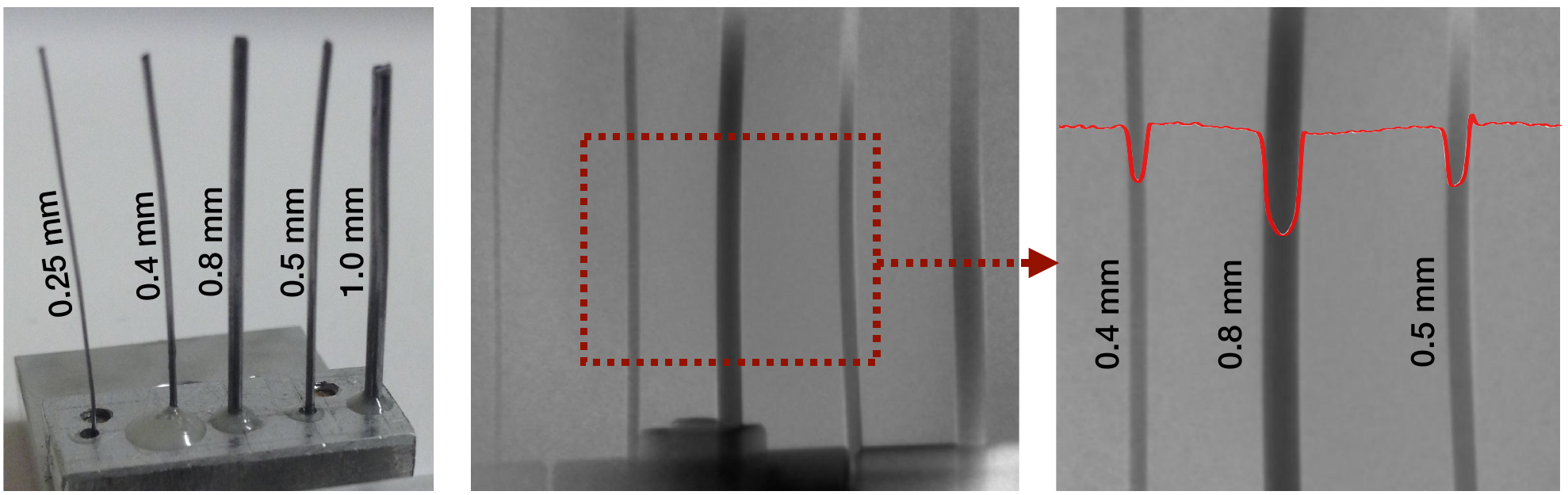}
   \caption{Tantalum wire array (left), its PRIOR radiograph with \unit[3.6]{GeV} protons (middle) and a zoomed central part of the middle image (right) with a transmission profile across the wires (red curve). \label{fig:wires}}
\end{figure}

In preparation for dynamic experiments (see Sections~\ref{sec:uewe} and \ref{sec:dynamic}), static mockups of the dynamic targets (exploding tantalum and copper wires) were radiographed to determine how the PRIOR magnifier could measure the density distribution inside an expanding metallic wire. A proton radiograph of a tantalum wire array is shown in Fig.~\ref{fig:wires}. The central wire with $800\,\rm \mu m$ diameter is a mockup of the un-exploded wire for the first dynamic experiments. The transmission profiles across the wires with different diameter show that there is sufficient sensitivity and resolution to measure the areal density of the wire while it expands.

The transmission~-- target thickness dependency, $T(z)$ of a proton radiography system can be described by a simple analytic model to a percent accuracy~\cite{mor:2011}:
\begin{subequations}
    \begin{align}
        &T(z) = \frac{e^{\displaystyle-z / \lambda}}{T(0)} \left(1 - e^{\displaystyle-\frac{\theta_{c}^2}{2\,(\theta^2(z) + \phi^2)}} \right), \label{eq:dens_a}\\
        &\theta(z) = \frac{13.6\,\mathrm{MeV}}{\beta p c}\sqrt{\frac{z}{X_o}}\left[ 1 + 0.088\log_{10}\!\left(\frac{z}{X_o}\right)\right] . \label{eq:dens_b}
    \end{align}
    \label{eq:dens}
\end{subequations}

The first term $e^{\displaystyle -z / \lambda}$ in Eq.~\eqref{eq:dens_a} describes the removal of the protons due to nuclear interactions in the target material with the nuclear collision length $\lambda.$ The second term is due to the multiple  Coulomb scattering: $\theta(z)$ is the RMS scattering angle and $\theta_c$ is the angular acceptance of the magnifier defined by its collimator. Assuming a normal distribution of the multiple scattering, the $\theta(z)$ dependency can be approximated with sufficient accuracy by the Moli\'ere theory~\cite{lyn:1991}, Eq.~\eqref{eq:dens_b}. Here $p$ is the proton momentum, $\beta$ is the proton velocity in units of the velocity of light $c,$ and $X_o$ is the radiation length of the material. The only empirical parameter in the model Eq.~\eqref{eq:dens} is the angular spread of the beam $\phi$ due to the beam emittance and overburden material (e.~g. vacuum windows or air) downstream of the target which is added in quadrature to the multiple scattering angle $\theta(z)$ caused by the object, and describes a small attenuation of the beam when there is no object.

\begin{figure}[ht!]
   \includegraphics[width=\columnwidth]{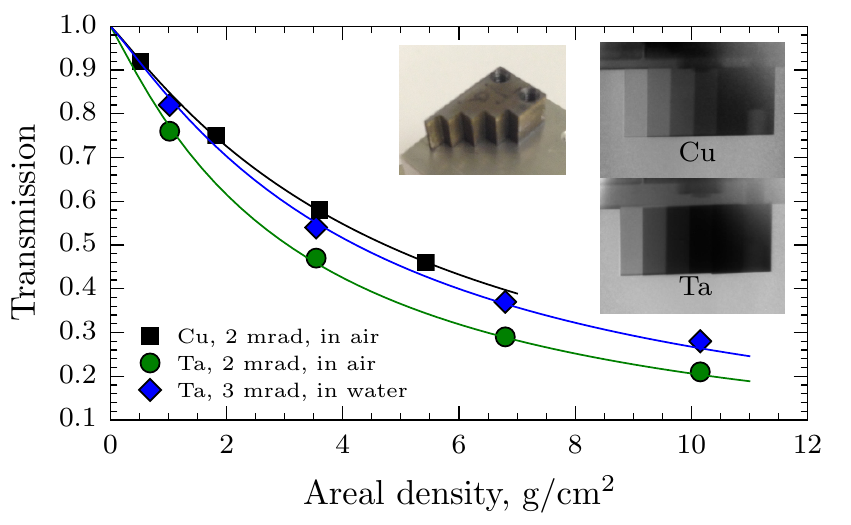}
   \caption{Transmission~-- areal density calibration with step wedges and wires (Fig.~\ref{fig:wires}). The tantalum step wedge target is shown in the left inset and the proton radiographs of the identical copper and tantalum step wedges ($0.56,$ $2.06,$ $4.07$ and $6.05\,\rm{mm}$ step thicknesses) are shown in the right insets. The transmission data measured with different collimator acceptance angles $\theta_c$ for the targets placed in air and in water (\unit[30]{mm} thickness, see sec.~\ref{sec:uewe}) is shown along with the model Eq.~\eqref{eq:dens} (solid lines) which has only one fitting parameter~--- the angular spread of the beam, $\phi$ (see explanations in the text). \label{fig:dens}}
\end{figure}

To test the density sensitivity of the microscope and to obtain the transmission~-- target thickness calibration $T(z)$ needed for dynamic experiments, radiographs of a series of step wedge targets were collected (see Fig.~\ref{fig:dens}). Identical tantalum and copper step wedges were used. In order to replicate the conditions of the dynamic commissioning, the step wedge targets were placed in the middle of the dynamic explosion chamber (see sec.~\ref{sec:uewe}, Fig.~\ref{fig:UEWEscheme}) which was filled with water or left in air. Two collimators with $\theta_c$ equal to \unit[2 and 3]{mrad} were used. Fig.~\ref{fig:dens} shows that the measured transmission is in good agreement with the simple analytic model Eq.~\eqref{eq:dens}. The data demonstrates a remarkable density sensitivity and proves that the PRIOR microscope can be used for radiographic density measurements.

\begin{figure}[ht!]
   \includegraphics[width=\columnwidth]{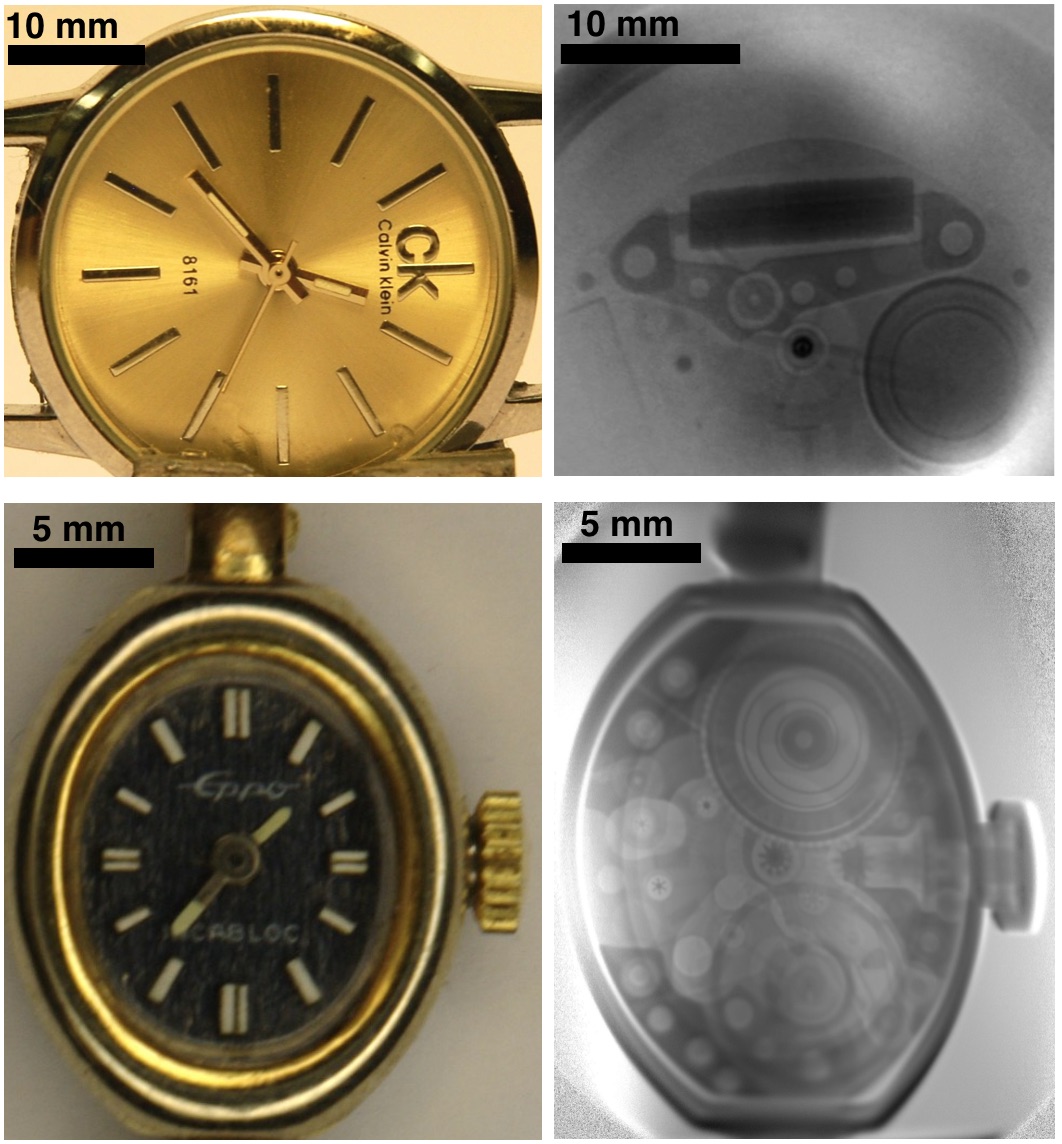}
   \caption{PRIOR \unit[3.6]{GeV} proton radiographs of complex targets. Top: a quartz watch and a radiograph of its central part. One can clearly see the battery and movement. Having a sufficient contrast, one can also see the hour, minute and even second hands of the watch. Bottom: a tiny mechanical watch. Despite a thick stainless steel case back, the fine details of the interior of the watch are well resolved: the crown and the mainspring, pivots and wheels, jewels, etc. \label{fig:watch}}
\end{figure}

Experience with proton radiography at the pRad facility at LANL has proven that radiographing various static objects is a good way to test a new radiography configuration. Extracting the geometry of challenging objects can often test the ability to resolve detailed structure of the objects, providing an opportunity to study and image beyond rolled edges, step wedges and other calibration targets. To fill this role a few \enquote{common} objects were radiographed for this effort, and two of them are shown in Fig.~\ref{fig:watch}: small quartz and mechanical watches. The slight non-flatness of the radiographic background is due to the data averaging and shot-to-shot beam position fluctuations. The obtained proton radiographs of these objects with complex interior structures clearly demonstrate the remarkable radiographic capabilities of the PRIOR setup.

As an unfortunate result of the first PRIOR run, we have observed a continuous degradation of the image quality and spatial resolution towards the end of the experiment. This phenomenon was attributed to the radiation damage of the PMQ lenses due to large fluences of spallation neutrons which are mainly produced in the tungsten beam collimator located in a close proximity to the third magnet (Fig.~\ref{fig:magdraw}, p.~\pageref{fig:magdraw}) as well as due to the primary protons scattered to large angles in the target and in the collimator. A significant radiation damage of neodymium-iron-boron PMQs has been also observed at LANL~\cite{dan:2014}. Because of this the 3D fields maps of all the magnets were measured after the first commissioning run. The results of the field distribution measurements have demonstrated a significant damage of the PMQs, especially of the first and the third lenses: the quadrupole strengths were reduced by \unit[10 -- 13]{\%} and the high-order field harmonics (relative sextupole, octupole and duodecapole field components) raised to the \unit[1.5 -- 2.5]{\%} level. This explains the degradation of the imaging performance of the system. We have also performed additional simulations and measurements of the PMQ radiation damage phenomenon~\cite{sch:2013,lan:2015,sch:2015} which confirm the results of the LANL study~\cite{dan:2014}. Unfortunately, the time between the static and dynamic PRIOR commissioning runs was not sufficient for re-magnetizing, reassembling and readjusting the lenses and we had to use the microscope in the dynamic experiments (see sec.~\ref{sec:dynamic}) in the same suboptimal state.

\section{Underwater electrical wire explosion}
\label{sec:uewe}

\begin{figure}[ht!]
   \includegraphics[width=0.9\columnwidth]{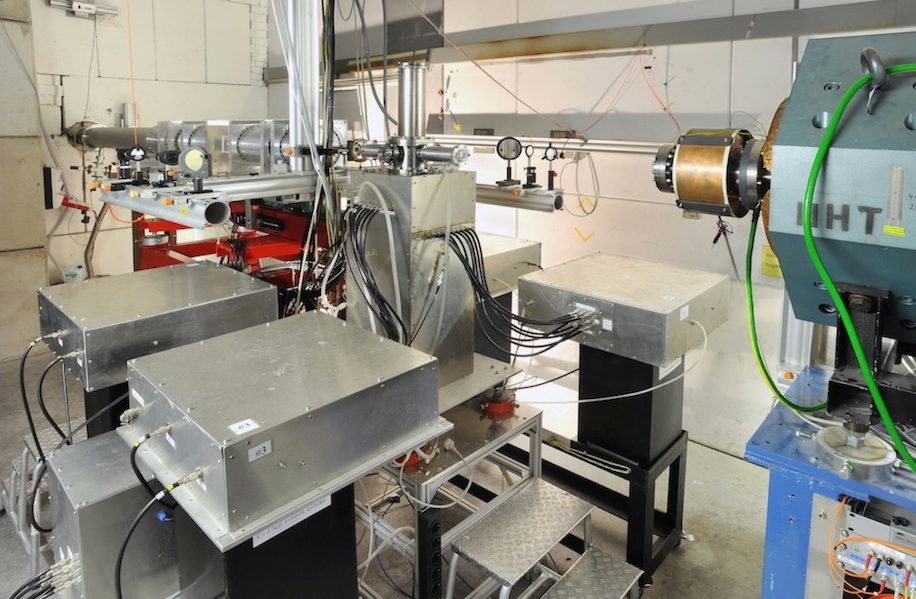}
   \caption{Pulsed power setup for underwater electric wire explosion experiments (UEWE) installed at the HHT area of GSI. A water filled UEWE explosion target chamber surrounded by four pulsed power generators is placed in front of the PRIOR proton microscope.   \label{fig:UEWEphoto}}
\end{figure}

Underwater electrical wire explosion (UEWE) is an efficient method for creating and studying warm dense matter in the laboratory~\cite{des:1998,she:2010,she:2012a}. The main advantages of the UEWE are the absence of the parasitic plasma formation along the wire surface due to high electric breakdown threshold (\unit[$>$300]{kV/cm}) of the water and relatively small wire expansion velocity ($10^5\,\rm cm/s$). These features allow to retain high current densities in the wire (up to $10^9\,\rm A/cm^2$) and therefore, by using a moderate pulsed power generator, one can create dense strongly coupled plasmas characterized by \unit[10 -- 100]{kJ/g} specific energy, near-solid density and several eV temperature. The main challenge in warm dense matter experiments is the determination of plasma parameters, and especially temporally and spatially resolved measurements of the target density. High energy proton microscopy is a unique diagnostic technique to address this problem.

\begin{figure}[ht!]
   \includegraphics[width=\columnwidth]{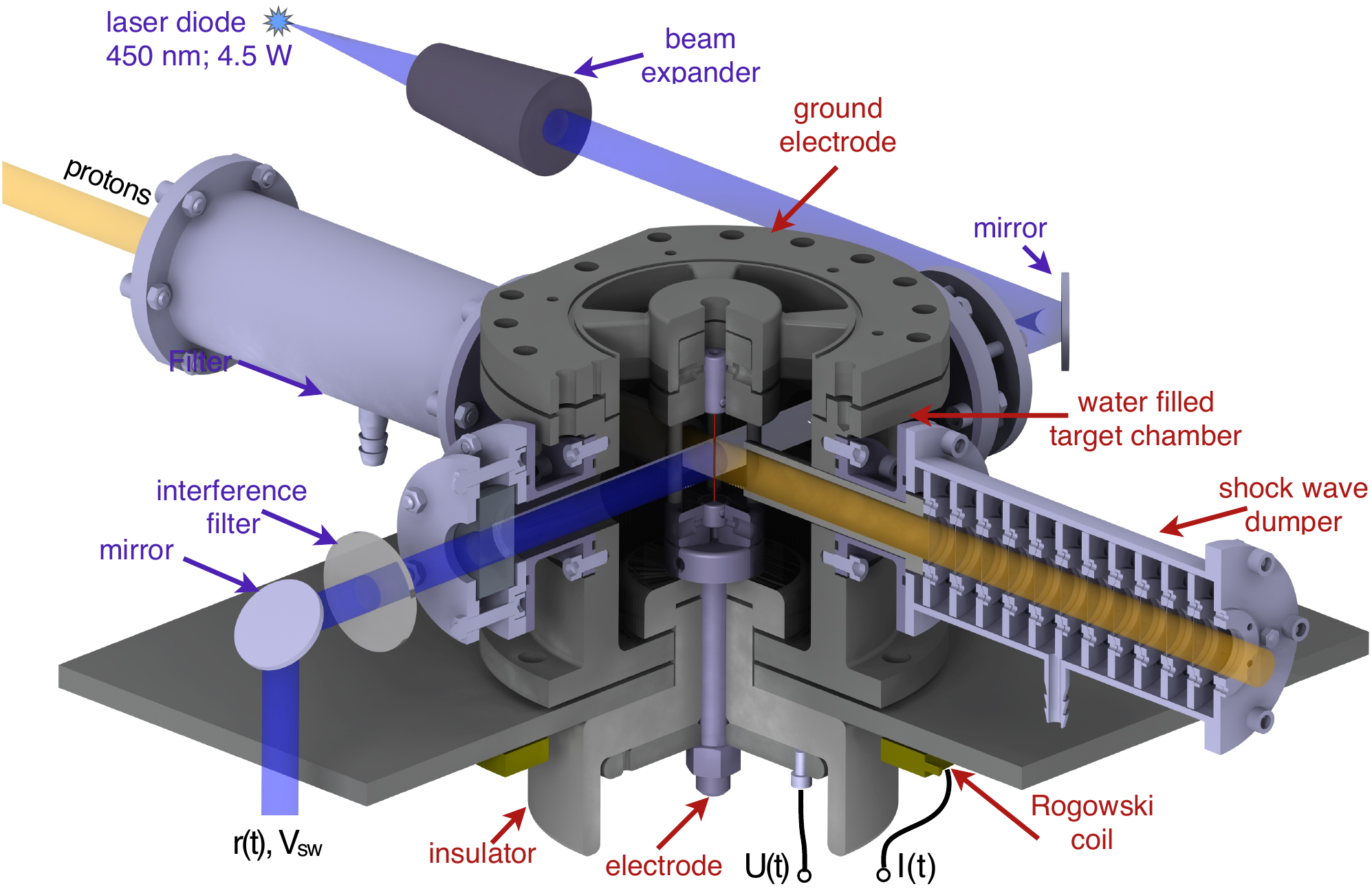}
   \caption{Scheme of the UEWE explosion chamber and target diagnostics. A thin exploding wire (red) in the center of the chamber is illuminated from one side by a proton beam for HEPM measurements, and from the other side~--- by a laser diode backlighter for optical diagnostics. \label{fig:UEWEscheme}}
\end{figure}

A new pulsed power UEWE setup has been constructed for dynamic commissioning of the PRIOR microscope (Fig.~\ref{fig:UEWEphoto}). The pulsed power generator (\unit[10]{$\mu\rm F$}, up to \unit[50]{kV} charging voltage and \unit[12.5]{kJ} stored energy) consists of four modules and can drive currents of about \unit[200]{kA} in amplitude and $1.8\,\rm\mu s$ rise time through a load at charging voltage of \unit[35 -- 40]{kV.}

During the PRIOR experiments, tantalum wires (\unit[0.8]{mm} diameter and \unit[40 -- 50]{mm} length) were quickly heated by a pulsed current to dense plasma conditions characterized by specific enthalpy level about \unit[5 -- 15]{kJ/g} and $\sim\!\rm{km/s}$ expansion velocities. Tantalum has been chosen for the experiments due to its high density which allows for a higher contrast of proton radiographs.

The construction of the UEWE explosion chamber and the scheme of the target diagnostics is shown in Fig.~\ref{fig:UEWEscheme}. A wire is placed in the middle of the \unit[11]{cm} diameter stainless steel explosion chamber which is filled with deionized water. A special effort has been taken to design water shock dampers in order to minimize the amount of material needed to separate the water-filled UEWE explosion chamber and the vacuum PRIOR beam line. The dampers (see Fig.~\ref{fig:UEWEscheme}) are \unit[18.5]{cm}-long, \unit[6.6]{cm}-diameter aluminum pipes containing from nine to twelve $150\,\mu\rm m$-thick Mylar foil stacks in holders with \unit[22]{mm} opening. A shock wave induced in water by an exploding wire consequently breaks the Mylar foils filling the damper pipe with water until it is completely stopped before the vacuum window of the beam line. From the explosion chamber side, the dampers are equipped with plastic insets sealed by a thin rubber. The insets allowed the reduction of the water layer thickness in the proton beam direction down to \unit[30]{mm}.

\begin{figure}[ht!]
   \includegraphics[width=\columnwidth]{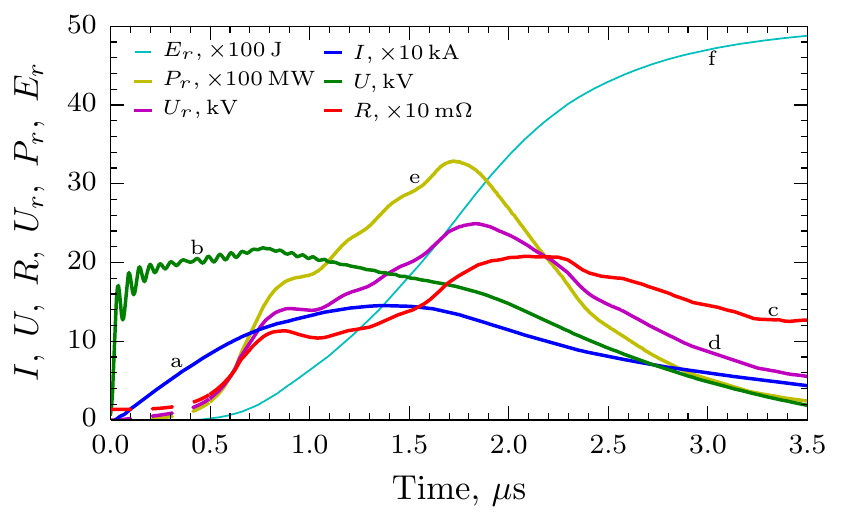}
   \caption{Electrical measurements in the UEWE experiment. Measured waveforms of the current $I$ (a) and voltage $U$ (b) are shown along with wire resistance $R$ (c), resistive voltage $U_r = I\cdot R$ (d), deposited power $P_r = I^2 \cdot R$ (e) and energy $E_r = \int_0^t P_r(\tau) d\tau$ (f). \label{fig:rcl}}
\end{figure}

The current flowing through a wire $I(t)$ and the voltage drop on the load $U(t)$ were measured by a Rogowski coil and a resistive voltage divider, respectively. The resistance of the wire $R(t)$ can be obtained from the following equation:
\begin{equation}
    U(t) = I(t) \cdot R(t) + L(t)\frac{dI(t)}{dt} + I(t)\frac{dL(t)}{dt}.
    \label{eq:rcl}
\end{equation}

The last term in Eq.~\eqref{eq:rcl} gives a small correction at a \unit[5]{\%} level which can be applied if the wire radius $r(t)$ is known from, e.g., optical or radiographic measurements: $dL/dt \approx \mu_0 \ell / 2\pi \cdot r^\prime(t) / r(t),$ where $\ell$ is the length of the wire and $r^\prime(t)$ is its radial expansion velocity. Eq.~\eqref{eq:rcl} can also be used in its integral form (equation for the energy balance), and under the assumption of a constant load inductance $L(t) = L_0,$ the resistance of the exploding wire $R(t)$ can be obtained by solving the following equation:
\begin{equation}
    \int_0^t I^2(\tau)R(\tau)d\tau = \int_0^t I(\tau)U(\tau)d\tau - \frac{1}{2}L_0I^2(t).
    \label{eq_rcle}
\end{equation}

This integral equation was solved by forward substitution with subsequent smoothing of the result. The value of the load inductance $L_0$ was adjusted while solving Eq.~\eqref{eq_rcle} so that the resistivity of the wire in the beginning of the discharge is equal to the known resistivity of solid tantalum~\cite{gat:1983}. The results of electrical measurements of a typical UEWE experiment with PRIOR is shown in Fig.~\ref{fig:rcl}. The plateau on the resistance signal at about $0.8\,\rm\mu s$ corresponding to the measured enthalpy of \unit[0.5 -- 1]{kJ/g} can be attributed to the melting transition and the following rapid heating of expanding liquid tantalum. The quick rise of the resistance at $1.1 - 1.4\,\rm\mu s$ and an enthalpy in the target material of about \unit[4]{kJ/g} may indicate the onset of a rapid evaporation.

In addition to the HEPM measurements with the PRIOR proton microscope (sec.~\ref{sec:dynamic}) and electrical measurements, an optical setup has been installed for target diagnostics. The optical diagnostics of the exploding wires (backlighting and thermal emission imaging) allows to determine the radius of the discharge channel as well as the velocity of the shock wave induced in water. The setup consists of a 450-nm CWL, 4-W fiber-coupled laser diode backlighter, two fast intensified CCD cameras (PCO DICAM PRO), a streak camera (HAMAMATSU C10910) and a set of lenses, mirrors, filters and beam splitters (see Fig.~\ref{fig:UEWEscheme}).

\begin{figure}[ht!]
   \includegraphics[width=\columnwidth]{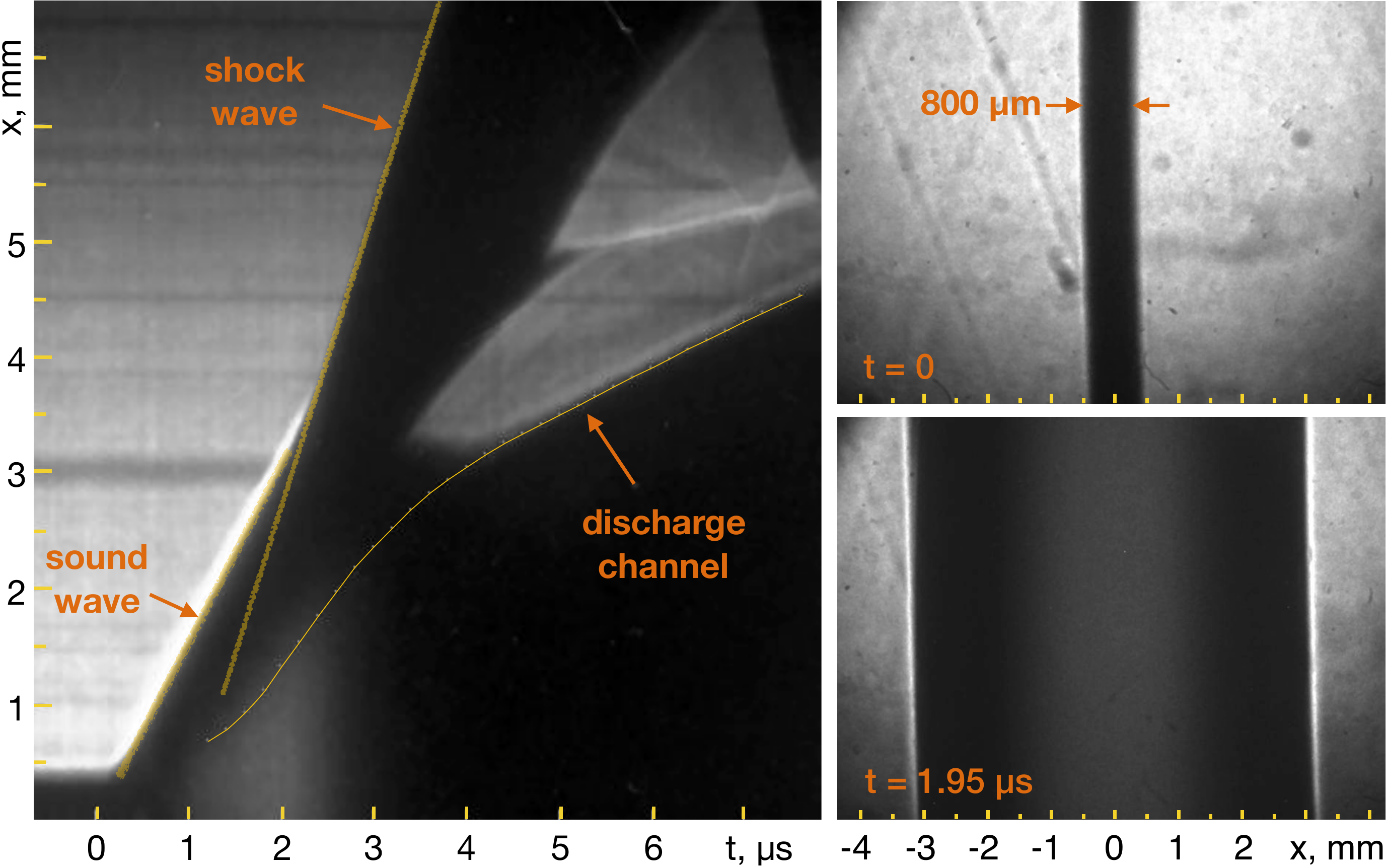}
   \caption{Optical diagnostics of the UEWE experiment (see explanations in the text). Left: streak camera record of exploding tantalum wire. Right: CCD camera images of the same wire at $t=0$ (top) and $t = 1.95\,\rm\mu  s$ (bottom). \label{fig:optics}}
\end{figure}

A typical streak camera record of an exploding tantalum wire with initial radius of \unit[0.4]{mm} is shown in Fig.~\ref{fig:optics}, left. Shortly after the beginning of the discharge ($t = 0$) a sound wave ($v \approx 1.5\,\rm km/s$) is launched in water by the rapidly heated and expanding wire. At the moment close to the wire explosion (fast boiling), a shock wave ($ v \approx 2.3\,\rm km/s$ corresponding to the pressure $P \approx 9.3\,\rm kbar$) is also launched and the thermal self-emission of the tantalum plasma becomes visible. At later times $t > 4\,\rm \mu s,$ the discharge channel radius can be traced again by the shadowgraphy. The shock wave and the wire discharge channel were also clearly visible in the CCD images (Fig.~\ref{fig:optics}, right).

\section{Dynamic experiments}
\label{sec:dynamic}
After the static beam time commissioning of the PRIOR setup and off-line tests of the UEWE setup, an experimental campaign of dynamic experiments with the PRIOR microscope took place at GSI. In comparison with the previous run, the proton beam intensity was increased by more than two orders of magnitude (up to $10^{11}$ protons per pulse) and a new beam diagnostics for high energy protons (scintillator screens and cameras) were integrated into the HHT beam line to ensure proper beam alignment and matching. Unfortunately, the shot-to-shot variations of the beam position and intensity distribution which were observed during the commissioning run with static experiments remained. The \unit[3.6]{GeV} proton beam has been delivered in four $\approx 40\,\rm ns$ long bunches with about \unit[150]{ns} inter-bunch spacing. Since the PRIOR image collection system was equipped with one fast camera, only one out of four bunches was used for the dynamic experiments.

Before the experiments with dynamic objects, a series of tests to determine the achievable temporal resolution of the microscope was conducted using a plastic scintillator (BC-400, decay time \unit[2.4]{ns}) and an intensified CCD camera (PCO DICAM PRO, fast shutter down to \unit[3]{ns}) in the image collection system. It has been shown that with the available proton beam intensity, one can achieve a \unit[5 -- 10]{ns} temporal resolution without significant deterioration of the imaging properties by gating the detector while using one of the four bunches. Due to the shot-to-shot instability of the beam, a \unit[20]{ns} detector gate was used for the dynamic target experiments.

The dynamic PRIOR commissioning was carried out using the developed UEWE setup and \unit[0.8]{mm} diameter tantalum wires. In total, about twelve successful dynamic shots with the PRIOR setup were completed. In these shots we have varied the power deposited in the wires by changing the wire length and capacitor charging voltage as well as the timing of the proton radiographs.

\begin{figure}[ht!]
   \includegraphics[width=\columnwidth]{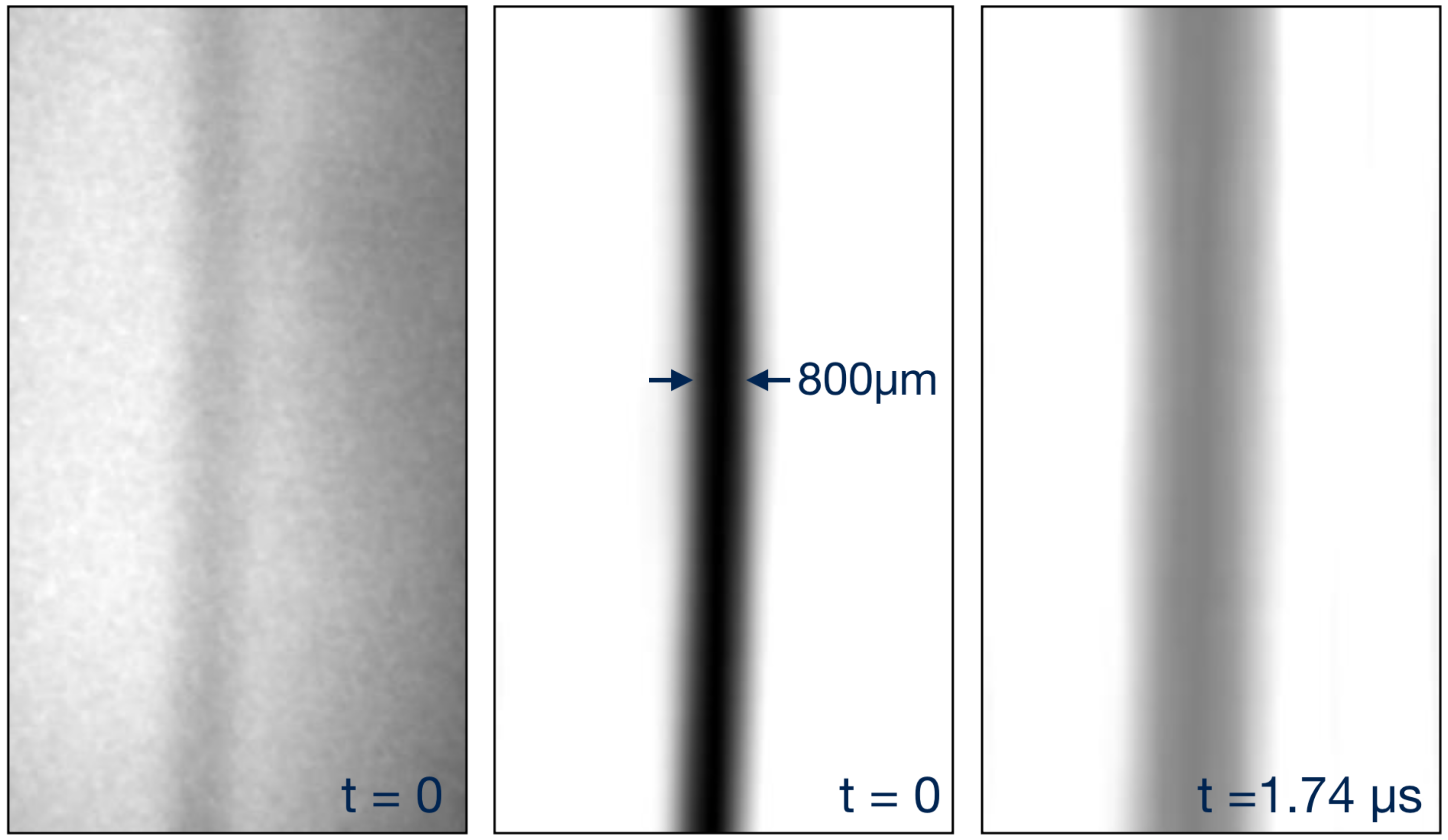}
   \caption{Processing of the dynamic proton radiographs. Left: raw radiographic image of a \enquote{static} (taken before shot) tantalum wire. Middle: areal density distribution in the static wire obtained after processing of the left image (see explanations in the text). Right: density distribution in the same wire during its explosion at $t=1.74\,\mu\rm s.$ All the radiographs were recorded with \unit[20]{ns} exposure. \label{fig:prad_wire}}
\end{figure}

Because of the shot-to-shot fluctuations of the beam position during the PRIOR dynamic commissioning run, the beam images could not be directly used for the processing of the dynamic radiographic data, and the information about the transverse beam intensity distribution had to be deduced from the dynamic target images themselves. The analysis of a large number of the recorded beam images has shown that the intensity distribution in the central area of the beam can be well approximated in each shot by an asymmetric Gaussian function. Using this knowledge, the transmission images were obtained by dividing the raw images by the empirical beam intensity distribution function with the function parameters fitted to the the same radiograph using the image areas which are not occupied by a target. An example of such data processing is shown in Fig.~\ref{fig:prad_wire} for \enquote{static} (taken before shot) and \enquote{dynamic} (taken during wire explosion) proton radiographs of a UEWE experiment.

\begin{figure}[ht!]
   \includegraphics[width=\columnwidth]{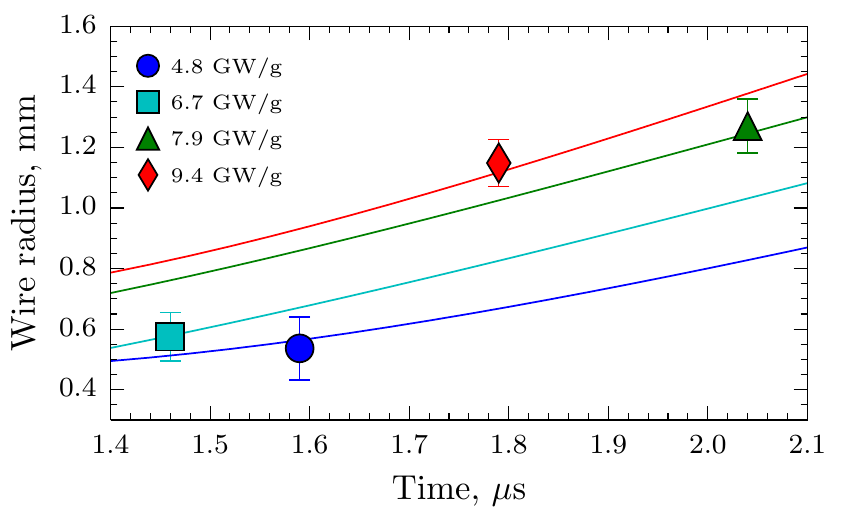}
   \caption{Radii of exploding tantalum wires measured with PRIOR proton microscope (symbols) and by optical streak diagnostics (solid lines, see also Fig.~\ref{fig:optics}, left). The initial radius of the wires was \unit[0.4]{mm}. The data is shown for four shots with different levels of the peak specific power deposition from 4.8 to \unit[9.4]{GW/g}. \label{fig:prior_res}}
\end{figure}

Although the ultimate goal of the UEWE experiments with the PRIOR microscope was to measure the radial density distribution of an exploding wire, the significant degradation of the PRIOR spatial resolution due to the radiation damage of the PMQ lenses as well as potential imperfections of the non-standard data processing procedure did not allow achieving this goal with a sufficient accuracy. Nonetheless, it was possible to use the obtained radiographic data for determining the radii of the exploding wires by deconvolving a Gaussian resolution blur and using the static wire radiographs as a reference. A comparison of these radiographic results with the results of the optical diagnostics (see sec.~\ref{sec:uewe}) is shown in Fig.~\ref{fig:prior_res} for four shots with different levels of the peak specific power deposition and correspondingly~--- different expansion velocities. The error bars of the radiographic results are caused mostly by the possible uncertainties due to the data processing procedure (deconvolution of the blur and the transmission data accuracy). Despite relatively large error bars, one can see that the obtained PRIOR proton radiographic results are in reasonable agreement with the optical measurements for all dynamic UEWE experiments.

\section{Summary and outlook}
\label{sec:summ}

As a result of a joint international effort, a new high energy proton microscopy facility has been designed, constructed and successfully commissioned at GSI. The beam time commissioning using intense \unit[3.5~-- 4.5]{GeV} proton beams delivered by the SIS-18 synchrotron has demonstrated $30\,\rm \mu m$ spatial and \unit[10]{ns} temporal resolutions with a remarkable density sensitivity. For the dynamic commissioning of the PRIOR magnifier, a new pulsed power setup for studying properties of matter under extremes has been developed and is operational at GSI.

The experiments have indicated that neodymium-iron-boron PMQ lenses are not an appropriate choice for a HEPM facility with high-energy and high-intensity proton beams due to the severe radiation damage of the magnets. Although samarium-cobalt permanent magnets are known to be more radiation-tolerant~\cite{dan:2014}, they are also not the right choice for a long-term operation of the PRIOR facility at FAIR, where more than two orders of magnitude higher proton beam intensities are expected.

Therefore the final design of the PRIOR proton microscope which is called PRIOR-II employs small but strong and radiation-resistant electromagnets (\unit[60]{mm} aperture and \unit[1.3]{T} pole tip field). The design also assumes that the PRIOR-II setup can be first fielded at the HHT area of GSI to use up to \unit[4]{GeV} protons delivered by the SIS-18 synchrotron for static or dynamic experiments, and later it will be transferred without modifications to a new experimental area at FAIR to use intense \unit[2 -- 5]{GeV} proton beams of the SIS-100 synchrotron. The new PRIOR-II facility will provide a magnification of about three at GSI and up to eight at FAIR due to a longer available drift length, with about $10\,\rm\mu m$ spatial resolution at the object.

\begin{acknowledgments}

The authors would like to thank the GSI's accelerator and technical teams for developing and timely integrating into the HHT beam line new beam diagnostics which was absolutely necessary for the commissioning of the PRIOR proton microscope (D.~Acker, M.~Bevcic, H.~Br\"auning, T.~Br\"uhne, E.~Dierssen, C.~Dorn, R.~Fischer, J.~J\"ohnke, R.~Haseitl, R.~Lonsing, A.~Petit, I.~Pschorn, C.~Schmidt, M.~Schwickert, K.~Steiner, R.~Vincelli and B.~Walasek-H\"ohne) as well as for the fine tuning of the GSI machines for the high energy and high intensity proton operation (W.~Barth, Y.~El-Hayek, B.~Franczak, C.~Omet,  D.~Ondreka and P.~Spiller). The authors would also like to thank K.~Gruzinskii and S.~Gleizer (Technion) for their kind assistance in designing and assembling of the UEWE pulsed-power setup. Work supported by GIF grant No. 1132-11.14/2011, BMBF grants No. 05K10RD1 and 05P12RDRBK and FRRC contract No. 29-11/13-17.

\end{acknowledgments}

\bibliography{prior}

\end{document}